\documentclass[]{spie}  

\usepackage{amsmath,amsfonts,amssymb}
\usepackage[all]{nowidow}
\usepackage{graphicx}
\usepackage[colorlinks=true, allcolors=blue]{hyperref}
\usepackage[scientific-notation=true]{siunitx}
\usepackage{subcaption}

\title{Model validation and tolerancing of scalar vortex masks in the High Contrast Imaging Testbed (HCIT) facility}

\author[a]{Niyati Desai}
\author[a]{Garreth Ruane}
\author[a]{Daniel Shanks}
\author[a]{Lorenzo K\"onig}
\author[a]{Susan Redmond}
\author[a]{Bertrand Mennesson}

\affil[a]{Jet Propulsion Laboratory, California Institute of Technology, Pasadena, CA 91109, USA}

\authorinfo{Send correspondence to ndesai@jpl.nasa.gov}

\begin{document} 
\maketitle

\begin{abstract}

The Habitable Worlds Observatory (HWO) mission will require coronagraphs capable of suppressing starlight at the $\sim 10^{-10}$ contrast level to directly image exo-Earths. High contrast achromatic coronagraphic masks are the missing critical component to achieving this. Vortex coronagraphs, particularly scalar vortex designs with an achromatic focal plane mask, offer key advantages. While all vortex coronagraph varieties provide high throughput, a small inner working angle, and rejection of low-order aberrations, the scalar approach enables dual-polarization observation in a single optical path. This simplifies instrument design and increases transmission by maintaining light from the planet in two orthogonal polarization states. In this work we test scalar vortex masks and investigate their contrast limitations. We perform phase metrology to assess the mask defects and manufacturing deviations and use it to refine the coronagraphic model used for electric field conjugation (EFC) algorithms and end-to-end simulations. We also measure the impact of model-mismatch with EFC by varying model parameters including clocking angle, and central wavelength in laboratory demonstrations. Finally, we validate our scalar vortex models against experimental results from the High Contrast Imaging Testbed (HCIT) facility at JPL by finding good agreement between lab and simulated performance. This ultimately helps to benchmark simulated contrast predictions for future scalar vortex coronagraph designs for HWO.
\end{abstract}


\section{INTRODUCTION}
\label{sec:intro}  

The Habitable Worlds Observatory (HWO) mission concept aims to directly image and spectrally characterize Earth-like exoplanets around nearby stars, requiring raw contrast levels on the order of $10^{-10}$ at small angular separations\cite{Astro2020_Report}. Achieving such performance depends on precise starlight suppression through a coronagraph system, which relies critically on both the design and fabrication of a high-performing focal plane mask and the accuracy of the optical model used in model-based wavefront control algorithms such as Electric Field Conjugation (EFC) \cite{GiveOn_2009,Desai_2024SCC}. As coronagraph mask designs become increasingly complex to mitigate chromatic limitations and fabrication constraints, it becomes essential to understand how mask and model accuracy affect coronagraph performance in the context of future missions like HWO.


\subsection{Scalar Vortex Coronagraphs}

Among the various coronagraph architectures under consideration for HWO, scalar vortex coronagraphs (SVCs) have emerged as a promising solution due to their low inner working angle, high throughput, and dual-polarization capability\cite{Ruane2019}. Unlike vector vortex coronagraphs (VVCs), which rely on the geometric phase and require additional polarization optics (e.g., waveplates, analyzers)\cite{Ruane2022}, SVCs manipulate the longitudinal phase and can operate identically on both orthogonal polarization states within a single optical path. The SVC applies a spiral phase pattern to the incoming wavefront, redirecting starlight outside of the Lyot stop and enabling high-contrast imaging of off-axis planetary light. 
This desired phase pattern is typically implemented by spatially varying the thickness or refractive index of a material. This method not only enables precise phase control but also supports dual-polarization operation—a key advantage over vector vortex designs. This dual-polarization capability simplifies system design and integration while preserving more of the planet signal. 

Recent work developing SVCs has focused on addressing the chromatic limitations of current mask designs to achieve deeper broadband contrast. New SVCs have introduced increasingly complex focal plane mask topographies—such as multi-zone or sawtooth phase patterns—optimized for achromatic performance \cite{Desai_2022,desai2023_Roddier, Desai_2024Roddier,Galicher2020}. While effective at mitigating chromatic leakage, these designs present new challenges in fabrication precision and model fidelity.

\subsection{Model-based Wavefront Control}

Modern high-contrast imaging systems rely on model-based wavefront control to suppress residual starlight after coronagraphic rejection. One of the most widely used techniques is Electric Field Conjugation (EFC), which iteratively adjusts deformable mirror (DM) commands to drive the intensity in the focal plane to zero based on a model of the optical system \cite{Malbet1995,Giveon2007}. EFC assumes that the model accurately predicts how changes in the DM shape affect the complex electric field at the final focal plane. Inaccuracies in the optical model—especially in regions of strong phase gradients introduced by the coronagraph—can lead to suboptimal DM solutions and degraded contrast performance, especially in broadband light where model errors accumulate across spectral channels.


The central motivation of this work is to investigate how specific types of model mismatch, particularly in the focal plane mask, affect the performance of SVCs when used with EFC in the lab. We aim to quantify the sensitivity of achievable contrast to mask alignment errors and fabrication deviations and to provide guidance on acceptable model tolerances for future missions such as HWO.

\section{MEASURING \& MODELING MASK DEFECTS}

\subsection{Metrology Efforts}

\begin{figure} [t]
    \centering
    \includegraphics[width=\linewidth]{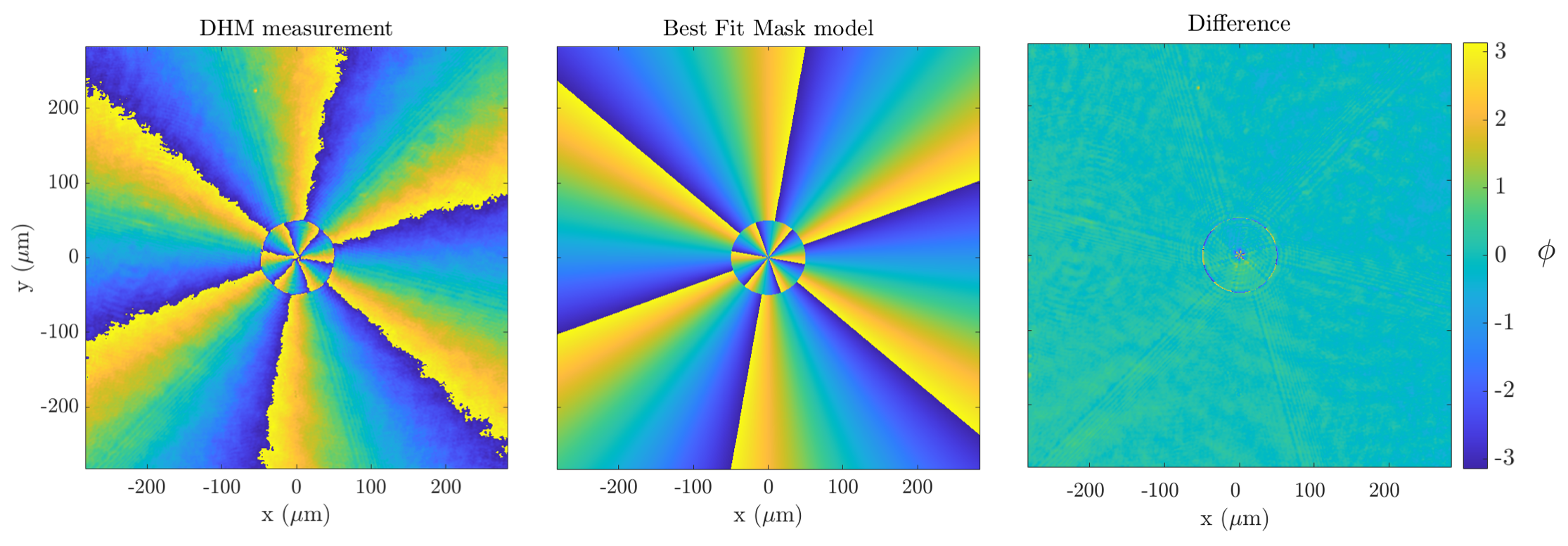}
    \caption{(left) Phase measurement of sawtooth with central phase dimple SVC prototype mask taken with the digital holographic microscope. (center) Best fit SVC model to DHM measurement. (right) Residual difference between measurement and model.}
    \label{fig:dhm_fit}
\end{figure} 


One approach to improving the model fidelity of complex structured coronagraph masks is through advanced mask metrology. This involves obtaining accurate phase measurements of the vortex masks, with particular attention to the regions near phase discontinuities. For VVCs, commonly used techniques include Mueller matrix measurements to quantify retardance and polarization-dependent leakage, as well as optical microscopy to identify point defects on the mask surface \cite{Llop-Sayson2024}. In the case of scalar vortex masks, optical microscopy remains useful for initial inspection, but since there is no polarization leakage, the phase must be measured in other ways. This could be done by using a profilometer to measure the surface profile and then translating these measurements to phase. Alternatively, phase measurements could be directly acquired interferometrically using either a digital holographic microscope (DHM) or a Zygo interferometer\cite{Bloemhof2003}.

This work measured and modeled a prototype of a combined sawtooth and central Roddier dimple scalar vortex mask presented by Desai et al. 2024a\cite{Desai_2024Roddier}. Further details regarding mask design and prototype fabrication can be found in Desai et al. 2024b\cite{Desai_2024SPIE}. Figure~\ref{fig:dhm_fit} shows how a DHM phase measurement of the SVC (left plot of Figure~\ref{fig:dhm_fit}) is used to identify defects and constrain the model parameters that best fit the prototype. The vortex model (center plot of Figure~\ref{fig:dhm_fit}) can then be subtracted from the phase measurement to reveal surface smoothness and any fabrication defects. The plot on the right of Figure~\ref{fig:dhm_fit} shows the residual phase difference and is indicative of the quality of mask fabrication. There is great agreement between the measured prototype and the model except at the central defect and around the radial phase discontinuities. 

While the DHM proved useful for qualitatively evaluating fabrication quality, the profilometer and the Zygo interferometer provide more consistent and reliable measurements\cite{Desai_2024SPIE}. Although these two methods show stronger agreement with each other than with the DHM results, a common limitation among these methods is their inability to resolve the central region of the mask, particularly within 10 microns of the optical axis. As a result, the precise structure near the center remain difficult to characterize. This limitation motivated Section~\ref{subsec:defect}, which investigates the performance impact of a worst-case central defect and potential strategies to mitigate its effects.

These ongoing efforts to refine mask metrology techniques are key to determining whether the surface quality achieved through grayscale etching of these prototypes is sufficient to support the starlight suppression levels required for high-contrast imaging with HWO. As future prototypes are developed with new manufacturing techniques and tested, these techniques will be essential to compare between them.

\subsection{Impact of the central defect}
\label{subsec:defect}


\begin{figure} [t]
    \centering
    \includegraphics[width=\linewidth]{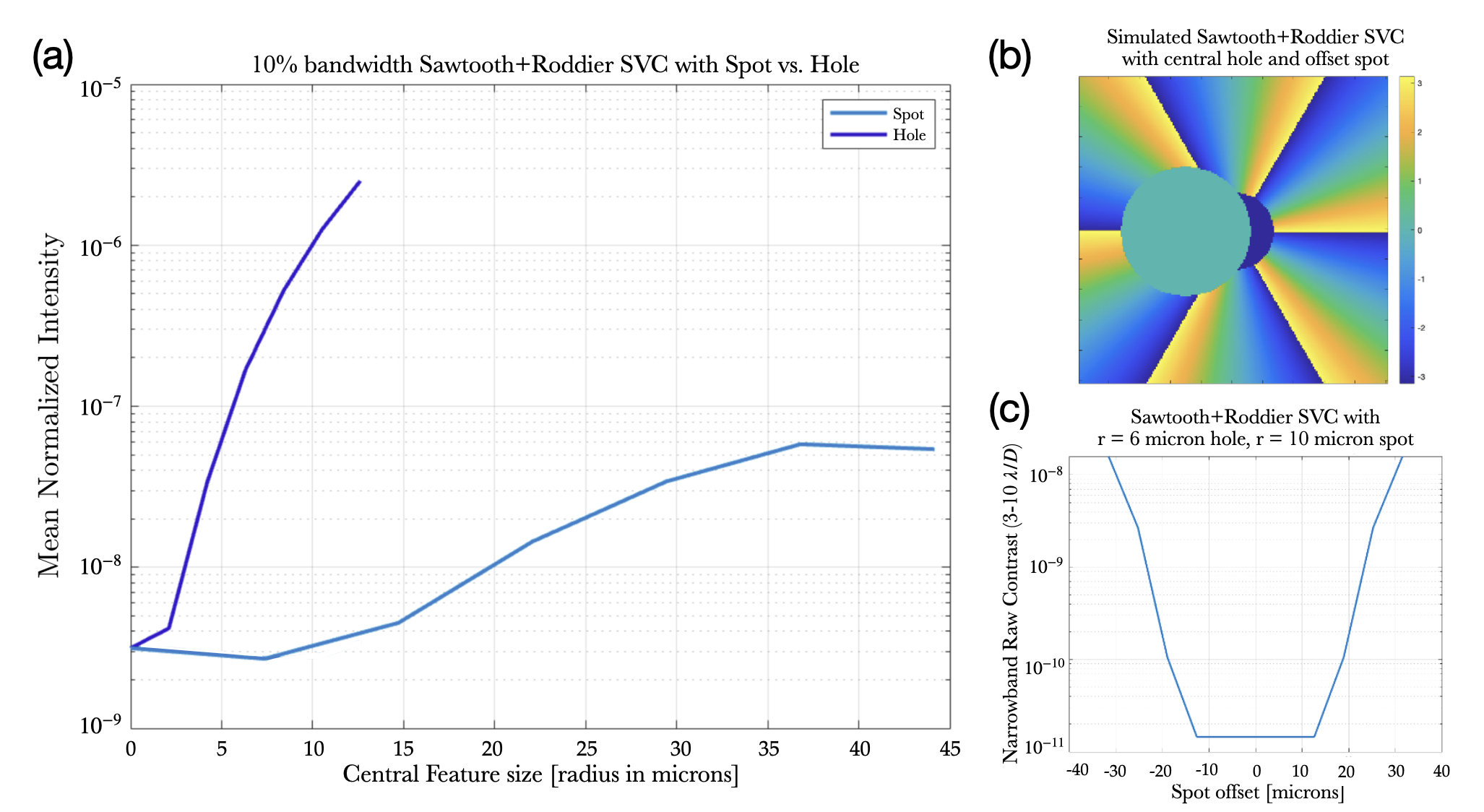}
    \caption{(a) Simulated broadband contrast of a sawtooth with Roddier dimple SVC with varying sizes of central defect hole or opaque spot. (b) Simulated central hole and and offset opaque spot. (c) Simulated narrowband contrast with various offsets of a r = 10 micron opaque spot covering a r = 6 micron central defect.}
    \label{fig:hole_v_spot}
\end{figure}

One of the most critical regions of a vortex phase mask during fabrication is its center. Since the mask is placed at the focal plane, the core of the starlight point spread function concentrates near the optical axis, causing any imperfections at the very center of the mask to significantly degrade performance. For scalar vortex masks fabricated using lithographic techniques, the center presents a particular challenge. Fabrication resolution limits can make it difficult to produce the precise phase pattern near the center, where the required azimuthal phase ramp becomes singular.

To quantify the impact of a central defect, we performed simulations where the center of the vortex mask was replaced with a small region of constant phase—effectively modeling a fabrication-limited defect. We then varied the size of this region to observe its effect on contrast performance. The results, shown by the purple curve in Figure~\ref{fig:hole_v_spot}a, reveal a rapid degradation in contrast as the defect diameter increases, with significantly worse contrast even for defects only a few microns wide.

A common mitigation strategy for vector vortex masks, is to deposit a small opaque spot over the central region to block the problematic light entirely \cite{Ruane2018}. We investigated a similar approach for scalar vortex masks by adding an opaque mask over the vortex center in our simulations and varying its diameter. As shown by the blue curve in Figure~\ref{fig:hole_v_spot}a, this method effectively recovers high contrast as long as the opaque spot remains smaller than approximately 20 microns in diameter. Beyond this size, the opaque spot still mitigates the impact of the central defect, but raises the contrast floor, limiting the vortex coronagraph’s ability to suppress starlight effectively.

In addition to spot size, we also modeled the impact of misalignment when depositing the opaque spot over the central defect. Figure~\ref{fig:hole_v_spot}b shows the geometry of this offset case, where a 10-micron radius spot is displaced laterally with respect to a 6-micron radius central defect. The corresponding contrast performance, shown in Figure~\ref{fig:hole_v_spot}c, demonstrates that the opaque spot is only effective if it completely covers the defect. When the spot is offset such that any portion of the defect is left uncovered, the leaked light dominates and significantly reduces the achievable contrast. This result underscores the importance of precise alignment during the fabrication or integration process, even when using mitigation strategies like opaque central masks.

\section{MODEL-MISMATCH EXPERIMENTS}

To evaluate the sensitivity of SVCs to model mismatch, we conducted a series of controlled experiments using the In-Air Coronagraph Testbed (IACT) at JPL\cite{Baxter2021}. We kept the experimental setup fixed: a sawtooth scalar vortex mask in IACT, and only modified the model used in the EFC algorithm. Specifically, we focused on how errors in modeling the angular orientation, "clocking", and central design wavelength of scalar vortex masks affect contrast performance when using EFC. 

\subsection{Impact of clocking mismatch}

\begin{figure} [t]
\centering
\begin{subfigure}{.37\textwidth}
  \centering
  \includegraphics[width=\linewidth]{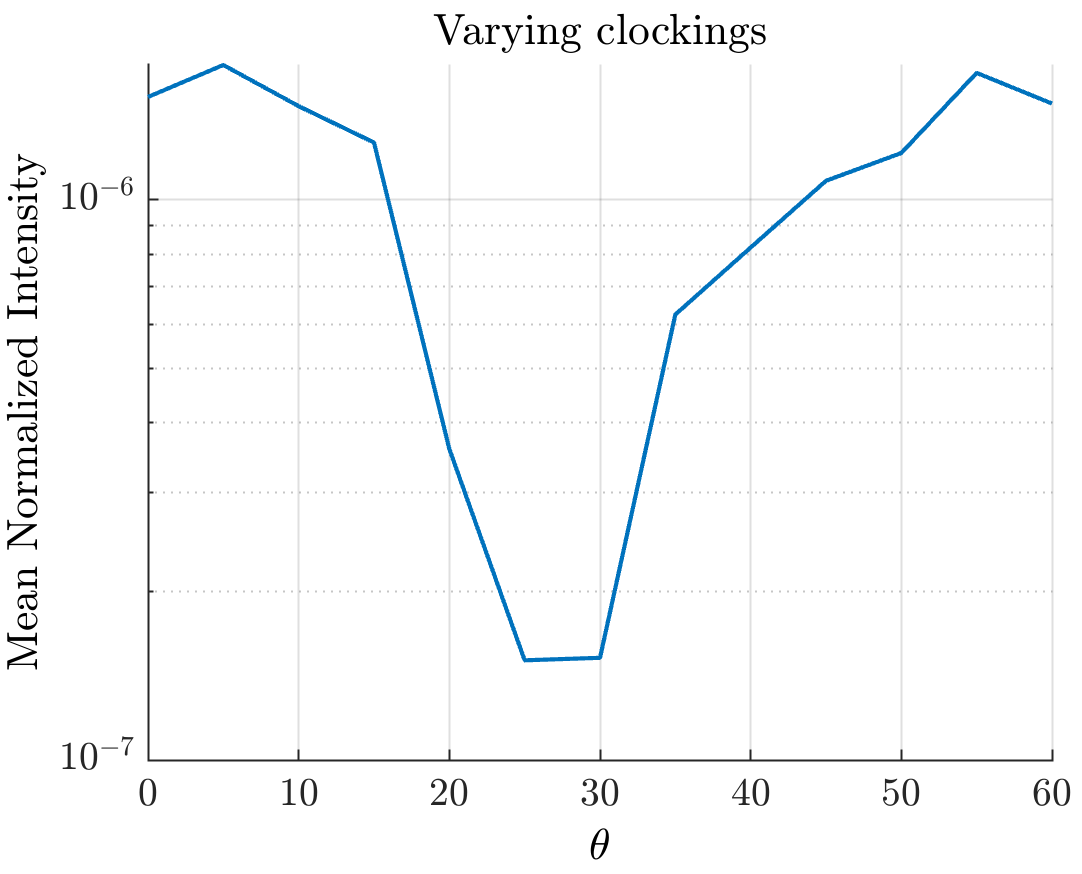}
  \caption{}
  \label{fig:clockingA}
\end{subfigure}%
\begin{subfigure}{.63\textwidth}
  \centering
  \includegraphics[width=\linewidth]{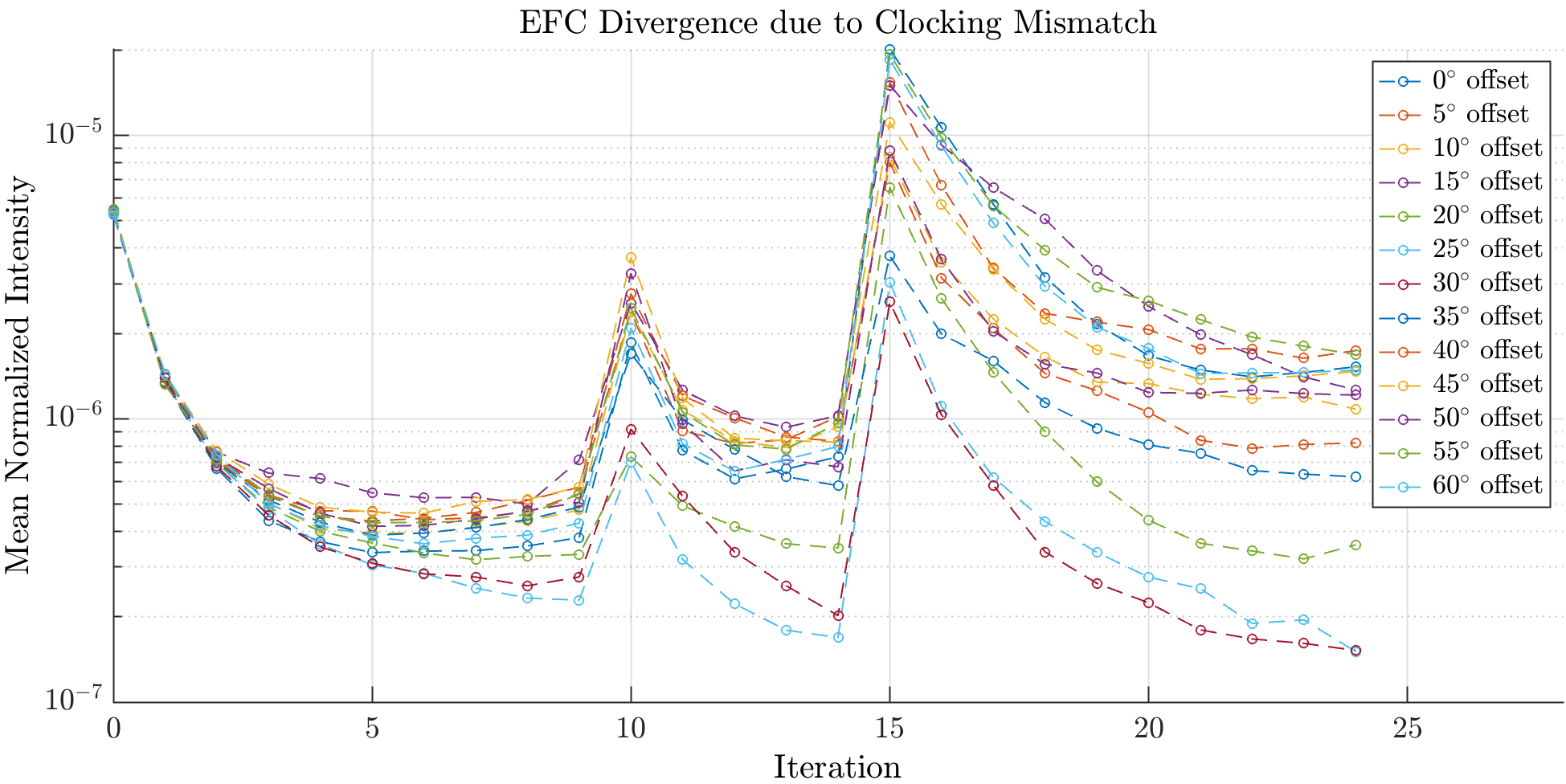}
  \caption{}
  \label{fig:clockingB}
\end{subfigure}
\caption{Sensitivity of broadband EFC performance to focal plane mask clocking mismatch. (a) Final contrast vs. modeled clocking angle, showing a clear optimum near 27°. (b) Contrast vs. EFC iteration for selected model angles, showing degraded convergence for mismatched clocking.}
\label{fig:clocking}
\end{figure}

For scalar vortex masks with discrete phase discontinuities—such as those with a six-sector sawtooth topography—accurately modeling the clocking angle is crucial for achieving optimal contrast. While scalar vortex masks are polarization-insensitive and do not require alignment with external polarization optics (unlike vector vortex masks), their phase structure must still be correctly oriented in the model used by EFC. One open question was how sensitive EFC performance is to a model-mismatch of the clocking angle of the focal plane mask.

To characterize this effect, we fixed the physical clocking of a scalar vortex mask on IACT and estimated its orientation using a diffuser-assisted imaging technique. This method involves illuminating the mask with incoherent light and slightly offsetting it in the focal plane to reveal the six phase discontinuities— an effect linked to Gibbs phenomenon. We then ran a series of EFC trials, varying only the clocking angle in the model used by the control algorithm while keeping the physical mask fixed.  Initially, narrowband EFC runs (2\% bandwidth) showed minimal sensitivity to clocking mismatch, with all runs achieving similar contrast levels near \num{1e-8}, suggesting that for spectrally narrow conditions, the DM can compensate for the model mismatch effectively. However, when performing broadband EFC trials (10\% bandwidth), a strong dependence on model accuracy emerged. In Figure~\ref{fig:clockingA} (left), we plot the final contrast achieved as a function of model clocking angle. The curve shows a distinct minimum—indicating optimal performance—at a clocking angle near 27°, consistent with the value estimated from the diffuser image.

The effect of clocking mismatch is further illustrated in Figure~\ref{fig:clockingB} (right), where we plot the contrast as a function of EFC iteration for different model clocking angles. When the model clocking is incorrect, the DM initially improves the contrast but eventually diverges from the optimal solution, resulting in degradation of performance. This highlights the importance of accurately characterizing focal plane mask alignment and incorporating it into model-based wavefront control routines.

\subsection{Impact of wavelength mismatch}

\begin{figure} [t]
\centering
\begin{subfigure}{.37\textwidth}
  \centering
  \includegraphics[width=\linewidth]{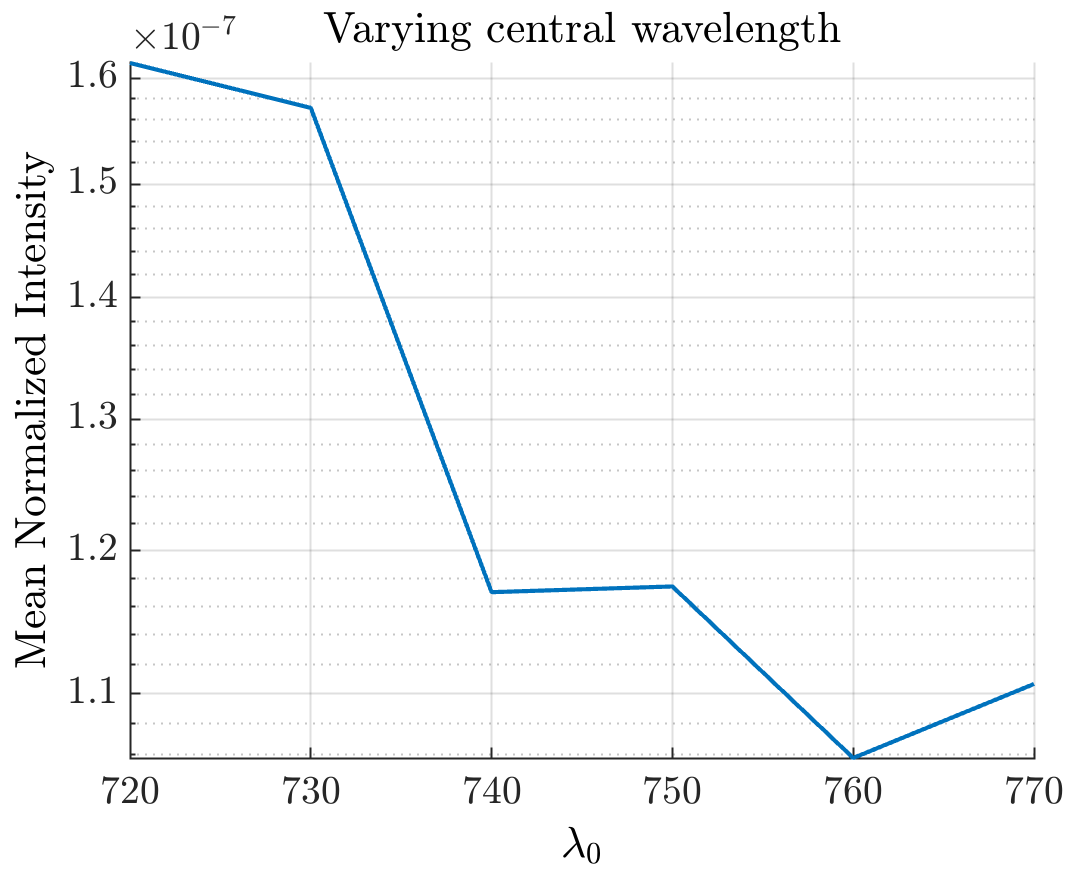}
  \caption{}
  \label{fig:wavA}
\end{subfigure}%
\begin{subfigure}{.63\textwidth}
  \centering 
  \includegraphics[width=\linewidth]{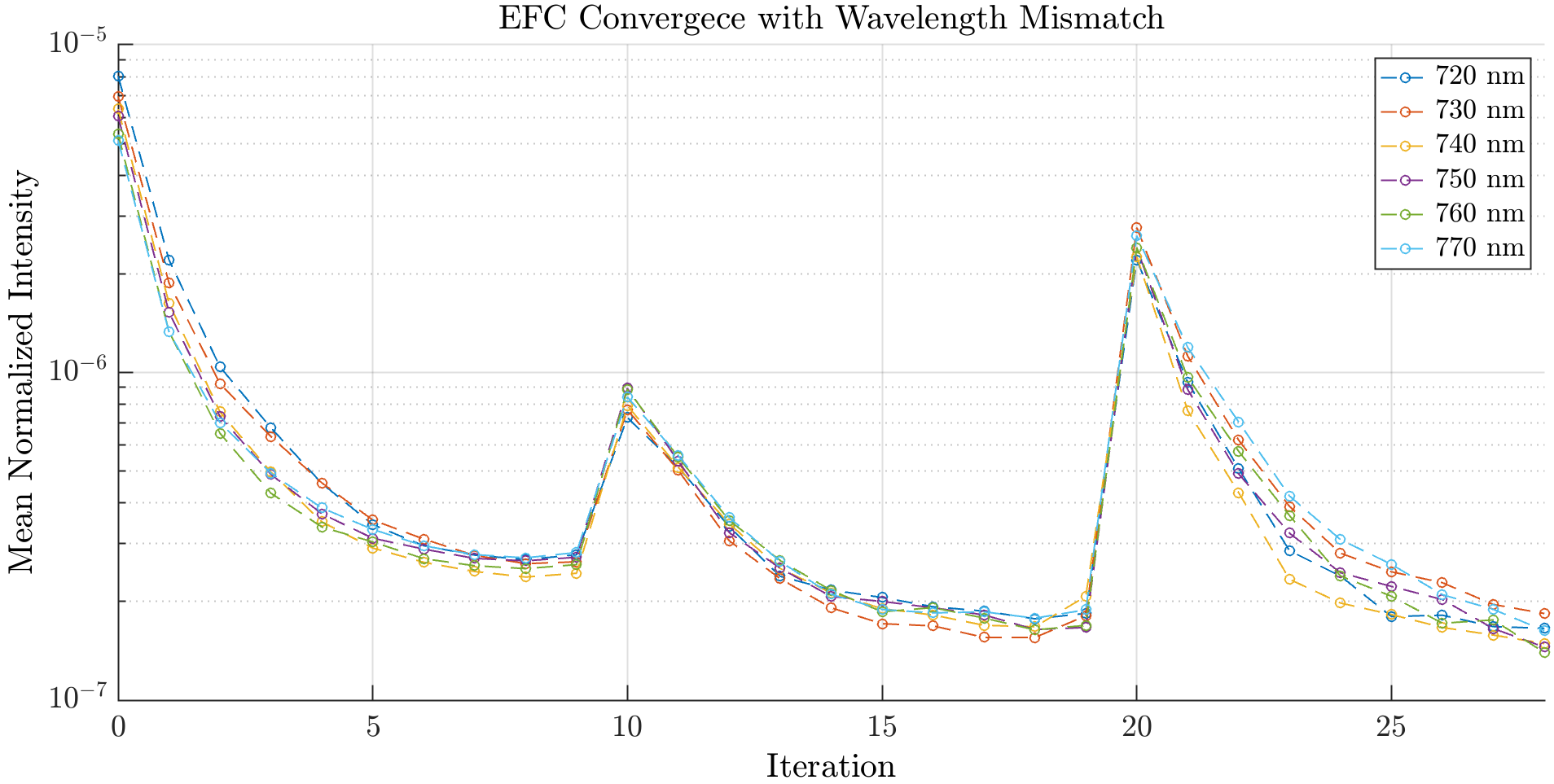}
  \caption{}
  \label{fig:wavB}
\end{subfigure}
\caption{Effect of central wavelength mismatch on broadband EFC performance. (a) Final contrast vs. modeled central wavelength, with optimal performance at 760 nm. (b) Contrast vs. EFC iteration for varying central wavelength $\lambda_0$, showing consistent convergence below \num{1.5e-7}. }
\label{fig:wav}
\end{figure}

In addition to investigating angular misalignment, we also aimed to understand how SVC performance is influenced by central wavelength mismatch between the model and the physical focal plane mask. This question is particularly relevant for broadband SVC performance, where small deviations in etch depth or material index during fabrication can effectively shift the phase response of the mask away from its designed center wavelength.

For scalar masks based on surface-relief structures, the phase imparted to the wavefront depends on the optical path difference across the etched features. Any fabrication error that alters the effective depth of the phase profile will still produce a scalar vortex—but with a shifted optimal wavelength $\lambda_0$. Because EFC relies on a model that assumes a specific $\lambda_0$, we expected that even a small offset could degrade performance, especially under broadband correction.

To explore this, we conducted a set of EFC runs on IACT using a scalar vortex mask nominally designed for 760 nm. In each experiment, we kept the hardware configuration fixed, but varied only the central wavelength assumed in the optical model used by EFC. The physical performance of the mask does not change across trials; only the wavelength-dependent modeling of the mask and the operating central wavelength are shifted.

The results of this experiment are shown in Figure \ref{fig:wav}. As expected, the best performance is achieved when the model matches the actual design wavelength of 760 nm (Figure \ref{fig:wavA}). However, unlike the clocking mismatch case, the degradation in performance is much less significant with wavelength mismatch(Figure \ref{fig:wavB}). Even with a ±20 nm offset in model wavelength, the contrast remains below \num{1.5e-7}, indicating that EFC is relatively robust to small $\lambda_0$ offsets in this range. This result shows that it is the combination of the coronagraph mask and the DM solution that needs to be achromatic for HWO contrast requirements.

\subsection{Leveraging Model-Free Approaches}

In situations where it is unclear whether the performance limitation arises from inaccuracies in the optical model or from the physical mask itself, we found model-free “dark hole digging” algorithms, such as implicit Electric Field Conjugation (iEFC)\cite{Haffert_2023}, particularly effective.

\begin{figure} [t]
    \centering
    \includegraphics[width=0.75\linewidth]{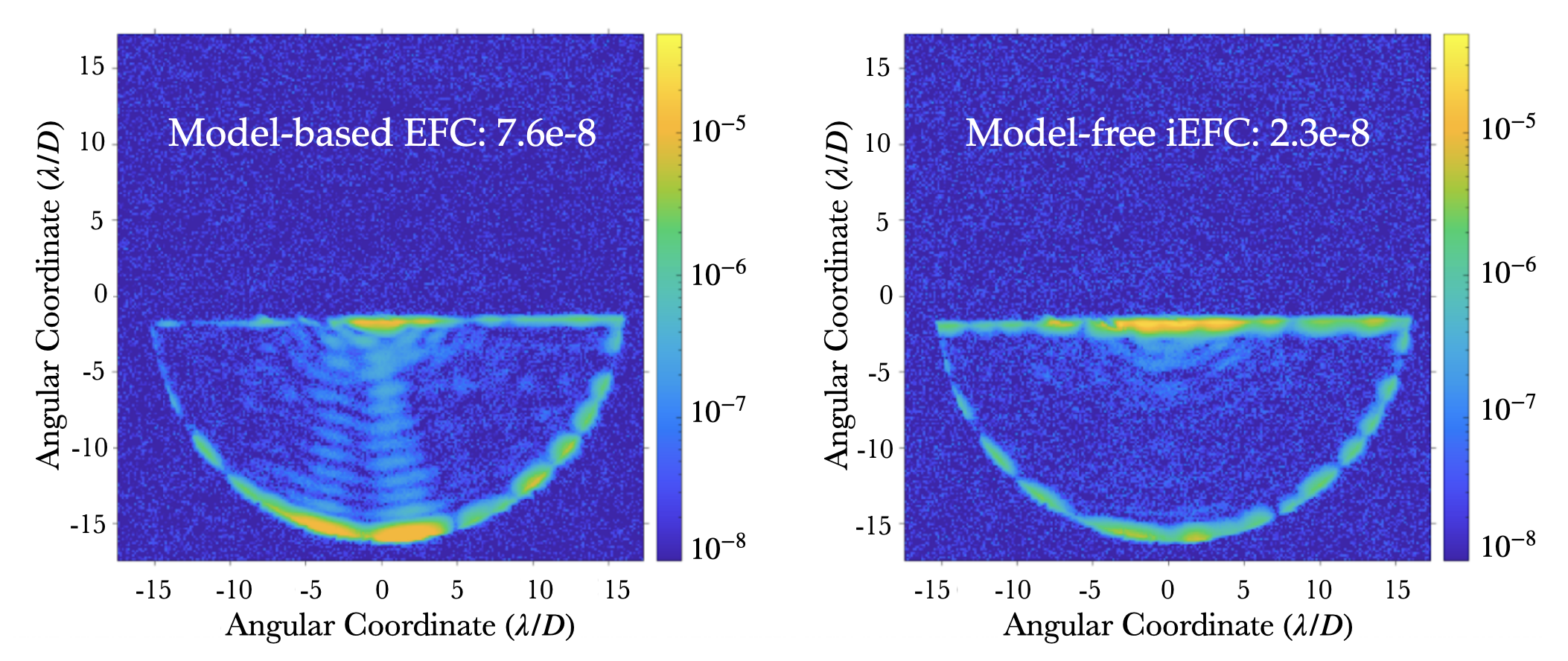}
    \caption{Side-by-side comparison of narrowband dark holes dug from 3-10 $\lambda/D$ on the In-Air Coronagraph Testbed at JPL with model-based EFC (left) and model-free iEFC (right) with a Lyot stop misalignment not matched in the model. }
    \label{fig:iefc}
\end{figure}

Although slower than model-based methods~\cite{Desai_2024SCC}, model-free approaches can achieve similar contrast when the system model is accurate. This makes them useful for validating models and identifying hardware alignment issues. In early model-based EFC tests, we observed persistent vertical streaks in the dark hole (shown in the left figure of Figure~\ref{fig:iefc}), suggesting a possible defect in the vortex mask, possibly at one of the six physical discontinuities. To investigate this, we ran iEFC, which optimizes DM commands without relying on a physical model. As shown in Figure~\ref{fig:iefc}, iEFC was able to remove the vertical streaks, confirming that the leakage could be corrected via appropriate DM actuation. 

This result indicated that the source of the leakage was not a fundamental limitation of the mask, but rather a mismatch in the system model. Further investigation revealed that the underlying issue was related to the Lyot stop alignment. During initial setup, alignment had been performed while the beam was still interacting with an inclined sector of the vortex mask, causing a lateral translation of the pupil image. This misalignment led to clipping at the Lyot stop, which was not captured in the original model. This demonstrates a good example of how ultimately tools like iEFC can be leveraged to ensure close model matching of the coronagraph system during instrument calibration and integration before deployment to space.


\subsection{Matching Experiment to Simulation}

\begin{figure} [t]
    \centering
    \includegraphics[width=0.75\linewidth]{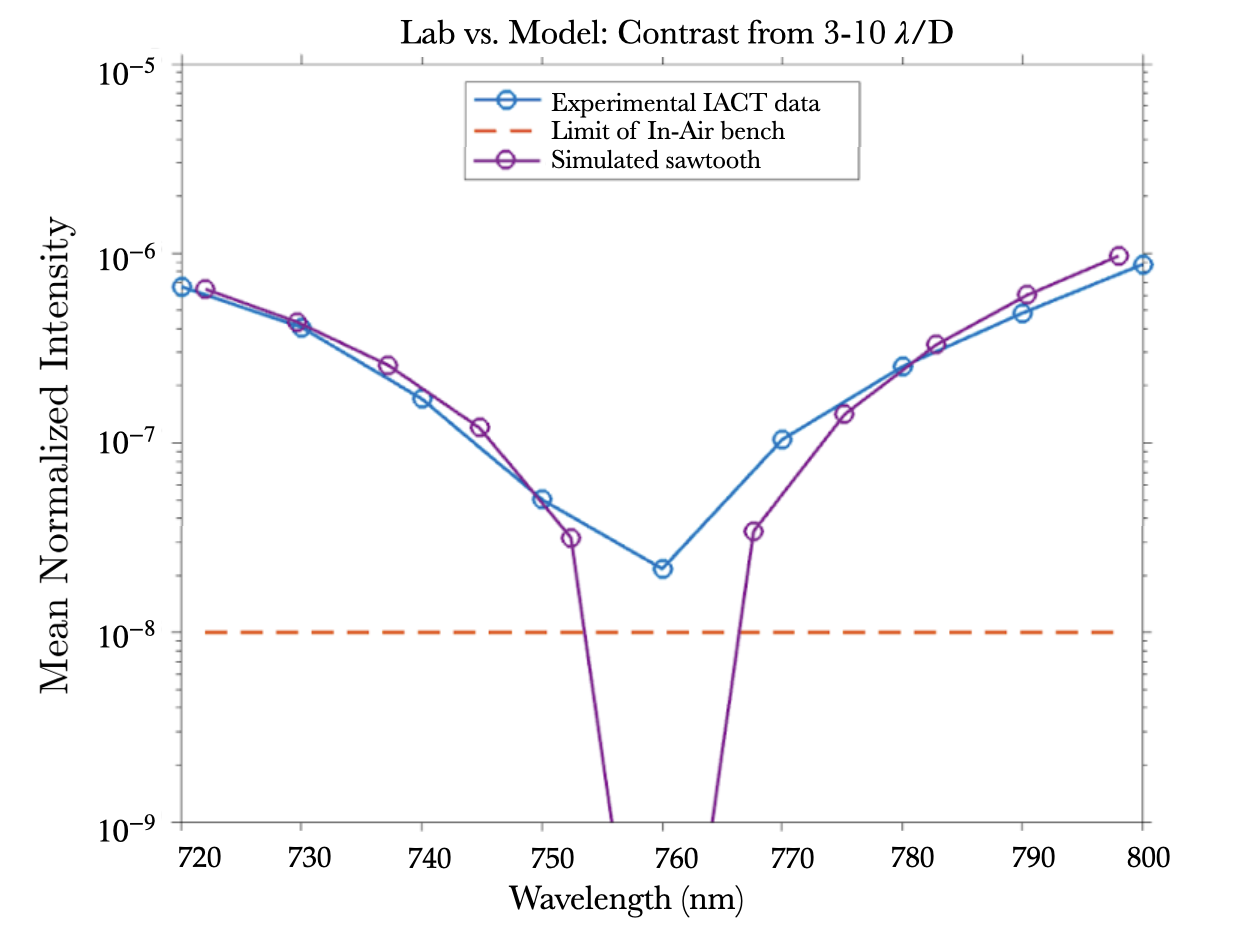}
    \caption{Comparison of experimental and simulated chromatic V-curve performance of a sawtooth scalar vortex mask. A very close agreement between IACT lab data and simulated performance can be seen.}
    \label{fig:lab_v_model}
\end{figure}

Crucially, Figure~\ref{fig:lab_v_model} shows a direct comparison between the experimental broadband contrast achieved with a sawtooth SVC on IACT (blue curve) and the contrast predicted from our refined optical model in FALCO \footnote{https://github.com/ajeldorado/falco-matlab} (purple curve). The excellent agreement between model and measurement validates the accuracy of our simulation tools and gives us strong confidence in our ability to predict the performance of future coronagraph designs. This validation is particularly important when exploring design trade-offs for new coronagraph masks, where predictive modeling plays a key role. It will also help to provide a deeper understanding of any experimental performance limitations we measure on IACT with future coronagraph prototypes. In particular, our next-generation SVC designs \cite{Desai_2024Roddier} already show predicted contrast levels below the current limits of the In-Air Coronagraph Testbed (shown by the orange dashed line in Figure~\ref{fig:lab_v_model}). The strong agreement between simulation and experiment demonstrated here motivates the transition of these and future prototype tests to vacuum testbeds, where the limitations of the bench no longer constrain performance.

\section{FUTURE PERSPECTIVES}
\label{sec:future}

In this work, we investigated the contrast performance of SVCs, with a focus on the sensitivity of these masks to both fabrication defects and modeling inaccuracies in the context of model-based wavefront control. Through detailed phase metrology, simulation studies, and experimental validation using IACT, we demonstrated how small mismatches in clocking, wavelength, and central fabrication defects can significantly affect contrast performance—particularly in broadband operation.

Simulations showed that the presence of even micron-scale central defects can degrade contrast by orders of magnitude, and that mitigation strategies such as opaque central spots are only effective if they are precisely aligned and properly sized. Furthermore, EFC performance was found to be highly sensitive to clocking angle mismatch, while more robust to small central wavelength offsets. These findings provide valuable insights into the required tolerances for future SVC mask prototype fabrication and model calibration.

Additionally, we compared the IACT experimental results to simulated contrast predictions for the same coronagraph architecture and observed excellent agreement across a broad wavelength range. This close alignment between lab and model results confirms the reliability of our models to predict performance for future coronagraph development. This is a critical step toward validating the true capabilities of these and future masks and advancing toward the $10^{-10}$ contrast levels required for HWO.

\acknowledgments 

We gratefully acknowledge our manufacturers at Zeiss for their excellent metrology and prototype fabrication, as well as their enthusiasm and collaboration in developing something new. We also thank PhD student Morgan Foley for generously designing and 3D printing custom holders for these prototypes with remarkable speed and precision. 

This research was carried out at the Jet Propulsion Laboratory, California Institute of Technology, under
a contract with the National Aeronautics and Space Administration (80NM0018D0004). N.D. is supported by an appointment to the NASA Postdoctoral Program at the Jet Propulsion Laboratory, California Institute of Technology, administered by Oak Ridge Associated Universities under contract with NASA.

\bibliography{report} 
\bibliographystyle{spiebib} 

\end{document}